\documentclass[printer]{tPHM2e}

\usepackage{graphicx}

\begin{document}
\doi{10.1080/1478643YYxxxxxxxx}

\markboth{Morineau \it{et. al.}}{Nanoconfined anisotropic molecular fluid}

\title{Structure and relaxation processes of an anisotropic molecular fluid\\confined into 1D nanochannels}

\author{R. Lefort, D. Morineau, R. Gu\'egan, A. Mor\'eac, C. Ecolivet, \thanks{$^\ast$Corresponding
author. Email: ronan.lefort@univ-rennes1.fr}$^\ast$\\\vspace{6pt} Groupe Mati\`ere Condens\'ee et Mat\'eriaux, CNRS-UMR 6626, Universit\'e de Rennes 1, F-35042 Rennes cedex, France \\\vspace{6pt}
and M. Guendouz\\\vspace{6pt} Laboratoire d'Optronique, FOTON, CNRS-UMR 6082, Universit\'e de Rennes 1, F-22302 Lannion cedex, France\\\vspace{6pt}
\received{\today}}

\maketitle

\begin{abstract}
Structural order parameters of a smectic liquid crystal confined into the columnar form of porous silicon are studied using neutron scattering and optical spectroscopic techniques. It is shown that both the translational and orientational anisotropic properties of the confined phase strongly couple to the one-dimensional character of the porous silicon matrix. The influence of this confinement induced anisotropic local structure on the molecular reorientations occuring in the picosecond timescale is discussed.\bigskip
\end{abstract}

\section{Introduction}
\label{intro}

One of the very early motivations for introducing confinement in experimental studies of liquids and glass formers was the general agreement that there might exist a typical lengthscale measuring the dynamical molecular cooperativity \cite{confit2003}. This quantity is suspected to increase with decreasing temperature, and to contribute to the observed anomalous properties of the structural relaxation such as super-arrhenian slowing-down in fragile compounds. Introducing a maximum size to the system by means of a porous matrix gives the opportunity to emphasize experimental evidence of those dynamical cooperative clusters. Up to now, most theoretical expectations and more or less direct experimental studies converge to state that the characteristic size of these cooperative rearranging regions (CRR) does not exceed a few molecular sizes. Confinement effects have been mostly scrutinized in porous materials presenting cavities of nanometric dimension. A large variety of materials has been used, often silicates within different geometries, either isotropic (fractal aerogels, interconnected vycors, aerosils or gelsils) or template-controlled (MCM-41, SBA-15). Several experimental difficulties are encountered in such studies due to the unavoidable powder averaging of the measured observables and to the huge surface to volume ratio that is reached. Indeed, the surface contribution to the free energy of the confined liquid cannot be ignored anymore, and the awaited ultimate finite-size effects can be masked by dominant interfacial solid-liquid interactions \cite{Gubbins-JCP-2003}. Experimental evidence of such surface contribution was reported on simple liquids like toluene \cite{Denistol}, as a stretching of the dynamical self-correlation functions toward longer times, and a non-vanishing static contribution within experimental resolution. These data were interpreted by a very large distribution of correlation times, starting from very slow reorientations imposed by a pinning at the pore surface, to bulk-like dynamics at the center. Such a broadening of molecular dynamics frequencies over several decades is supported by numerical simulations \cite{Denistol}. This onset of glass-like dynamics in simple liquids appears as an essential consequence of confinement, but is not the only one, for also the structure and the thermodynamical stable and metastable states are affected \cite{Christenson-2001}.
These interrelated effects remain partly misunderstood even for simple fluids. A large effort towards a better understanding of more complex confined fluids is timely, for they are introduced in an increasing number of innovative applications of nanosciences (lab-on-chip, nanofluidics...). Liquid crystals (LC) are test-systems in order to investigate fluids with an increasing degree on complexity. Their structure is characterized by anisotropic order parameters giving rise to typical sequences of nematic or smectic phases. However, their molecular nature allows one to confidently assign dynamical contributions in terms of rotations, translational diffusion or librations, accessible to a large number of techniques like dielectric spectroscopy \cite{sinhaaliev}, neutron quasielastic scattering or light scattering (photon correlation spectroscopy, optical Kerr effect \cite{FayerTg, Fayercomp}). Both these structural and dynamical aspects have been widely studied on bulk samples, and make LC's model systems particularly suited to experimental investigations of confinement effects on complex fluids. Like more simple liquids, the thermodynamics of confined LC's is also profoundly modified. One of the most striking effect is the instability of the translational smectic ordering with respect to the quenched disorder field introduced by an irregular solid/LC interface \cite{bellinisci, Lehenypre}. This leads to a very important depression of the melting temperature of the confined crystalline phase, and the suppression of the nematic to smectic continuous transition, replaced by a frustrated short-range (nanometric) smectic ordering \cite{regispre}. 

We report in this paper new experimental data depicting structural and dynamical features of a confined LC in a nanoporous material of very low dimensionality.

\section{Experimental Details}

\subsection{Sample}
Fully hydrogenated liquid crystal (LC) 4-octyl-4-cyanobiphenyl (8CB) was purchased from Sigma-Aldrich and used without further purification. Columnar porous silicon (pSi) matrices were obtained from a heavily doped (100) oriented silicon substrate by electrochemical etching \cite{fabripSi, Lehmann-MS-00}. A porosity of 60 \%, made of a parallel arrangement of not-connected channels (diameter $\approx$ 300 \r{A}) is obtained, aligned in a layer of thickness $\approx$ 30 $\mu$m. This very large aspect ratio induces a quasi one-dimensional character of the porous structure, preferentially aligned on the macroscopic lengthscale. Thermal oxidation of the structure can be further performed, leading to a transparent layer well suited to investigations by optical techniques. 8CB was confined into pSi by capillary wetting from the liquid phase under vapour pressure in a vaccuum chamber and at a temperature of 60\r{ }C, well above the N-I transition temperature. The excess of LC was then removed by squeezing the wafers between Whatman filtration papers. The complete filling of the pores was characterized by confocal microRaman spectroscopy \cite{quenselas}.

\subsection{Neutron Scattering}
Neutron diffraction experiments were performed on the G6.1 double-axis diffractometer (cold neutron guide) at the Laboratoire L\'eon Brillouin (LLB, CNRS/CEA, Saclay), with an incident wavelength of 4.71 \r{A}. Quasielastic incoherent neutron scattering experiments were carried out on the time-of-flight (TOF) spectrometer G6.2 at the LLB, with a resolution of 107 $\mu$eV FWHM. The covered momentum transfer (Q) range was about 0.4 to 1.9 \r{A}$^{-1}$. The temperature was controlled to better than 0.1 K over a range from 100 K to 340 K using a cryoloop. Numerical deconvolution of TOF spectra were carried out following standard procedures using the ``QENSH'' software provided by the LLB. For both diffraction and TOF experiments, eight filled porous silicon wafers were stacked parallel to each other in a cylindrical aluminium cell, representing a total amount of confined LC of about 20 mg. 

\subsection{Spectroscopic Ellipsometry}
\label{expellipso}
Ellipsometry experiments were carried out with a Horiba Jobin-Yvon UVISEL spectrometer, in the visible range ($\lambda$ from 410 to 830 nm). The measured ellipsometric angles $\Psi$ and $\Delta$ are related respectively to the amplitude ratio and to the phase difference between the complex reflexion coefficients ($r_{p}$ and $r_{s}$) of the sample  (resp. parallel or perpendicular to the incident plane). From these two angles, new quantities are defined according to :

\begin{eqnarray}
\label{eqn:one}
I_{s}=sin(2\Psi)sin(\Delta) ; I_{c}=sin(2\Psi)cos(\Delta)
\end{eqnarray}

These functions $I_{s}$ and $I_{c}$ were fitted versus the wavelength using the Horiba ``DELTAPSI2'' software provided with the apparatus. The sample was modelled as a porous silica layer top of a silicon substrate. The dielectric dispersions of silicon and silica were described using standard library data. The dielectric dispersion of the organic liquid crystal was modeled by a uniaxial anisotropic single oscillator. The porosity of the layer was described using a classical anisotropic Bruggeman model \cite{Bruggeman} mixing silica and void dispersions for the empty matrix, or silica and 8CB dispersions for the filled matrix.

\subsection{Brillouin Scattering}
\label{expbri}

Brillouin scattering was performed with a triple pass tandem of Fabry-Perot (Sandercock model) with a Kr$^{+}$ ion laser (Coherent) at a wavelength of 647 nm. Laser power was kept at a few tens of mW in order to avoid a too big heating due to absorption of the bulk silicon supporting the film. At too high counting rates a shutter obturates the photomultiplier inducing discontinuities in the spectra. Typical spectra were accumulated during several hours. Spectra were fitted by an apparatus function convoluted to damped oscillator profiles (Brillouin lines) and a Debye relaxor.

\section{Results and Discussion}
\label{resdis}

All results presented hereafter were obtained close to room temperature, in the smectic phase of 8CB (bulk or confined) \cite{Ocko-ZPB-86}. In this temperature range, the liquid crystal 8CB confined in pSi experiences smectic ordering that remains short-range (SRO, $\xi\approx150$ \r{A}) due to quenshed disorder effects imposed by solid-liquid interfacial interactions \cite{regispre, Lehmann-MS-00}. These effects destroy the bulk second order nematic to smectic phase transition. As a consequence, the Bragg peak associated to this smectic SRO and appearing at $2\theta_{B}\approx8.6$\r{ } (at a working wavelength of 4.7 \r{A}) is significantly broader than the experimental resolution. 

\subsection{One-dimensional ordering}
\label{1do}

\begin{figure}
\centerline{\includegraphics{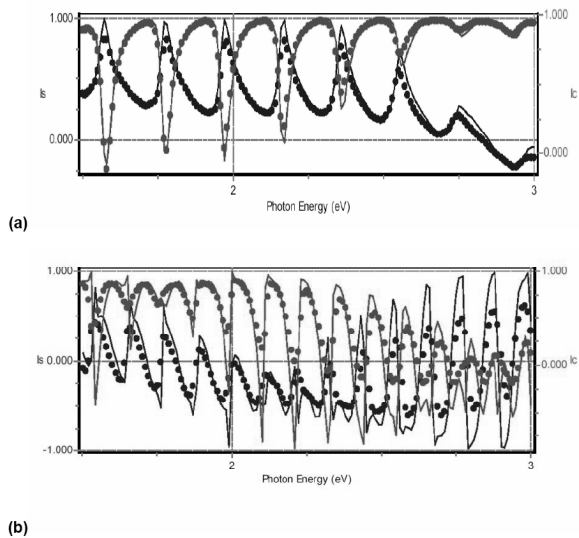}} \caption{Visible light ellipsometric spectra of (a) an empty porous silica layer, and (b) a porous silica layer filled with 8CB. Experimental data (filled circles) are compared to the fitted model (solid line).} \label{ellipso}
\end{figure}

Figure \ref{ellipso} displays ellipsometric spectra measured on empty porous silica and porous silica filled with 8CB. The interference fringes are due to multiple reflections at interfaces of the porous layer. It is therefore possible to deduce an estimation of the layer thickness, and typical values between 2.5 and 5 $\mu$m (depending on the sample) were found, in good agreement with independent SEM results. A simple comparison of both curves shows strong differences in the amplitude modulation of those fringes, as a consequence of the dispersion of the optical index of the confined liquid crystalline phase. This observation can be used as a direct characterization of a proper filling of the porous layer with 8CB. A semi-quantitative analysis was performed using the model described in section \ref{expellipso}. Best fits agree with an optical uniaxial birefringence of the filled porous layer of about $\Delta n\approx0.15$, with an optical axis aligned with the direction of the nanopores. The important differences in the ellipsometric data allow to separate unambiguously this additional birefrengence due to the liquid crystal from the natural optical anisotropy of the empty porous layer. These two contributions can be further more distinguished through the temperature dependence of the liquid crystal contribution. From the orientation of this measured optical anisotropy, it can be concluded that the orientational order parameter of the confined 8CB strongly couples to the one-dimensional character of the porous matrix.

\begin{figure}
\centerline{\includegraphics{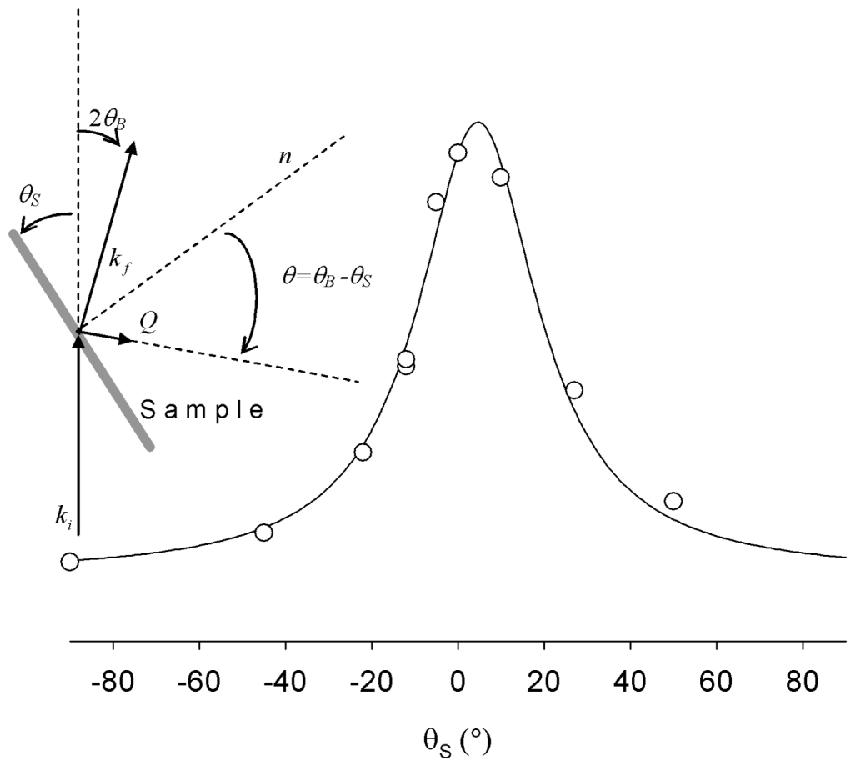}} \caption{Neutron cross section diffracted at the smectic Bragg position ($Q = 0.19$ \r{A}$^{-1}$) by 8CB confined into porous silicon. The inset sketches the experimental setup. The sample macroscopic orientation is defined by the angle $\theta_{S}$ between the silicon wafer plane and the incident beam.} \label{mosa}
\end{figure}

Figure \ref{mosa} shows the intensity of the Bragg peak associated to the smectic SRO measured at constant $q_{B}\approx0.19$ \r{A}$^{-1}$, while rotating the sample plane by an angle $\theta_{S}$ with respect to the neutron incident beam. The best fit using an arbitrary lorentzian form indicates that the maximum of the Bragg reflexion is peaked at $\theta_{0}\approx4.6$\r{ }, which corresponds to the sample orientation $\theta_{S}\approx\theta_{B}$ where the $Q$ vector is parallel to the nanochannels of the porous silicon layer. The linewidth is about 17\r{ }HWHM. Within this rather small mosaicity, the director of the smectic layers is obviously well aligned with the silicon channels axis.

\subsection{Rotational dynamics}
\label{rota}

Both ellipsometry and neutron diffraction results show that 8CB confined in porous silicon displays in the meantime orientational and translational order parameters that are macroscopically one-dimensionally oriented through a strong coupling with the peculiar low dimensionality of the matrix. Accordingly, the molecular dynamics of the confined phase might be also affected by this anisotropic ordering. Whereas translational diffusion should in some extent depend on the local properties of the smectic SRO domains, the rotational molecular relaxations should be more sensitive to orientational parameters. In this section, we report Brillouin and quasielastic neutron scattering results, that complementary probe the molecular relaxations in the picosecond time range.

\begin{figure}
\centerline{\includegraphics{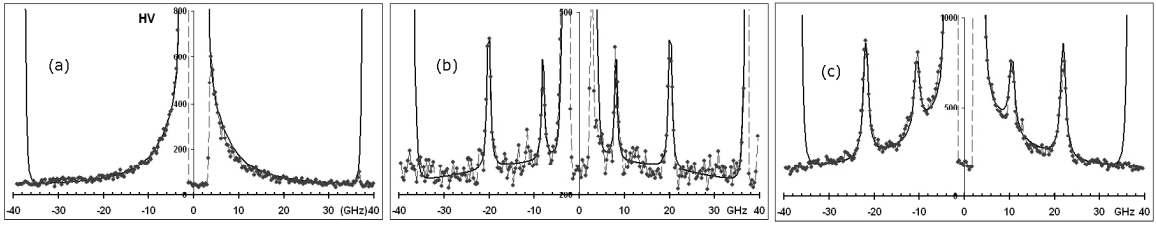}} \caption{Brillouin spectra measured on (a) bulk 8CB (b) empty porous silica, and (c) porous silica filled with 8CB.} \label{brillouin}
\end{figure}

\begin{figure}
\centerline{\includegraphics{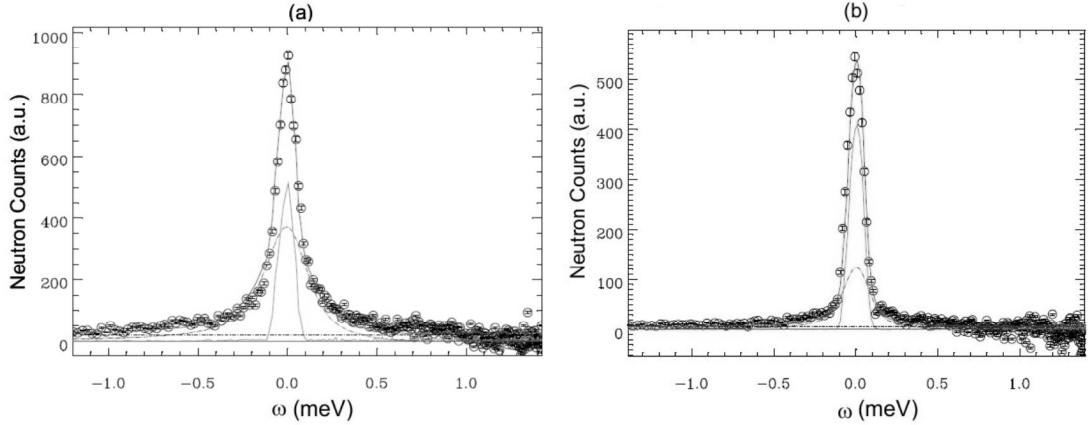}} \caption{Quasielastic spectra measured by time-of-flight neutron scattering at 295 K and $Q = 1.83$ \r{A}$^{-1}$, on (a) bulk 8CB, and (b) 8CB confined in porous silicon.} \label{mibemol}
\end{figure}

Figure \ref{brillouin} shows the comparison of Brillouin spectra obtained on porous silica filled with 8CB or on the separated constituents. The black line is the best fit according to the procedure detailed in section \ref{expbri}. The Brillouin doublets appearing on the spectra have different origins : the one at the lowest frequency is a longitudinal mode propagating tangentially to the surface, whereas at the highest frequency appears a longitudinal mode propagating mainly along the pores perpendicular to the surface. The response of the bulk liquid crystal consists in a single quasielastic lorentzian line of about 5 GHz HWHM (i.e. $\tau_{c}\approx30$ ps). The symmetry of the experimental setup suggests to assign this line to uniaxial rotation of the molecules around their long axis. The response of an empty porous silica matrix reveals strong Brillouin modes, sitting on a very broad quasielastic contribution (more than 35 GHz). Comparatively, the Brillouin spectrum of a filled porous layer can be described by a weighted linear combination of these individual responses, assuming a broadening of the acoustic modes. This damping is due to the coupling of the solid matrix to the viscosity of the liquid crystal (poroelasticity effect). A quasielastic contribution of about the same width as for the bulk 8CB can be seen on the data, suggesting that the uniaxial molecular rotations are only weakly affected by the confinement.
Quasielastic neutron scattering experiments were performed, in order to probe additional relaxations. Figure \ref{mibemol} shows the TOF spectra measured on bulk 8CB and 8CB/pSi. Spectra were fitted using one lorentzian plus one elastic (resolution limited) components. For bulk 8CB, the quasielastic contribution has a linewidth of about 0.25 meV ($\tau_{c}\approx3$ ps). Its occurence at large values of momentum transfer together with the presence of an elastic contribution would suggest that it also reflects localized reorientations (rotations, end-chain relaxations...). 

For 8CB confined in porous silicon, figure \ref{mibemol}(b) reports a significant increase of the elastic contribution, together with a decrease of the quasielastic linewidth down to 0.16 meV. This moderate slowing down of the fast reorientational motions agrees with the Brillouin results, and confirms that rotational molecular dynamics seem to be only weakly disturbed by the anisotropic confinement in porous silicon. On the other hand, the increase of the elastic contribution suggests that slow relaxations (slower than the TOF energy resolution) or long range translational diffusion might be much more affected. This is expected from a heterogeneous propagation of a surface induced slowing-down through the inner pore volume.

\section{Conclusion}
We report in this note new experimental results on a smectic liquid crystal confined into a low dimensional, macroscopically oriented porous silicon layer, which has been introduced recently \cite{regispre, quenselas}. The combination of a large panel of complementary techniques allows one to probe in parallel the structural and dynamical behavior of the confined molecules. It is shown that both orientational and translational order parameters of the confined liquid crystal couple to the anisotropy of the silicon matrix, and give rise to unusual structural properties reminiscent for some aspects of electric or magnetic field effects on bulk liquid crystals. The picosecond molecular dynamics of this oriented short-range ordered confined smectic phase were probed by light and neutron scattering techniques, suggesting that mainly large amplitude motions like long axis tumbling or long range translational diffusion, should be affected by anisotropic confinement effects.

\section{Acknowledgements}
The authors are thankful to J.-M. Zanotti and B. Frick, for expert advice on neutron scattering experiments and fruitful discussion, and also M. Stchakovsky for precious help on spectroscopic ellipsometry measurements.
%
%



\label{lastpage}

\end{document}